\documentstyle[prl, twocolumn, aps, epsfig]{revtex}
\begin{document}

\title{Grazing Collisions of Black Holes
via the Excision of Singularities}

\author
{
Steve~Brandt${}^{1}$,
Randall~Correll${}^{2,3}$,
Roberto~G\'omez${}^{4}$,
Mijan~Huq${}^{1}$,
Pablo~Laguna${}^{1}$,
Luis~Lehner${}^{2}$,
Pedro~Marronetti${}^{2}$,
Richard A.~Matzner${}^{2}$,
David~Neilsen${}^{2}$,
Jorge~Pullin${}^{1}$,
Erik~Schnetter${}^{1}$,
Deirdre~Shoemaker${}^{1}$ and
Jeffrey Winicour${}^{4}$
}
\address{
${}^{1}$Center for Gravitational Physics and Geometry,
Penn State University, University Park, PA 16802;\\
${}^{2}$Center for Relativity,
The University of Texas at Austin, Austin, TX 78712;\\
${}^{3}$National
Aeronautics and Space Administration, Washington, DC ~20546;\\
${}^{4}$Department of Physics and Astronomy,
University of Pittsburgh, Pittsburgh, PA 15260}

\maketitle

\begin{abstract} We present the first simulations of non-headon (grazing)
collisions of binary black holes in which the black hole
singularities have been excised from the computational domain.
Initially two equal mass black holes $m$  are
separated a distance $\approx10m$ and with impact parameter
$\approx2m$. Initial data are based on superposed, boosted
(velocity $\approx0.5c$)
solutions of single black holes in Kerr-Schild coordinates.
Both rotating and non-rotating black holes
are considered. The excised regions
containing the singularities are specified by following
the dynamics of apparent horizons.
Evolutions of up to $t \approx 35m$ are obtained in which
two initially separate apparent horizons are present for $t\approx3.8m$.
At that time a single enveloping apparent horizon forms,
indicating that the holes have merged. Apparent horizon area estimates
suggest gravitational radiation of about $2.6\%$ of the total mass.
The evolutions end after a moderate amount of time because of instabilities.
\end{abstract}

\bigskip

\noindent{{\bf Introduction:}
Gravitational wave detectors\cite{LIGO} will soon
begin searching for gravitational radiation from astrophysical binary
compact objects.  To understand these
observations, and to predict parameter regimes in which to search for
their radiation, efforts are underway
to model the interaction of compact sources.
We report here a direct numerical simulation of interacting spinning
\textit{black hole} binaries, in
genuinely hyperbolic (non-headon) trajectories.  The initial spin angular
momenta evolved here are either zero, or parallel to each other and perpendicular to the
orbital
plane.  The interior of the equal mass holes and their interior
singularities are excised from the computation. (Our method is {\it neither} restricted to equal
masses {\it nor} to parallel spins). Evolution is carried
out in a \textit{Cauchy} scheme, in which the state of the
gravitational system (the 3-spatial metric $g_{ab}$) 
and its rate of change (the 3-spatial extrinsic curvature $K_{ab}$)
are specified at one instant (i.e. on a 3-dimensional spacelike hypersurface) 
and are then
stepped to the next instant using an ``ADM"\cite{ADM} form of 
the Einstein evolution 
equations\cite{Italy}. 
The evolution is
unconstrained, and maintenance of the constraint functions with small
error is verified throughout the run.

This work extends previous work on headon 
encounters\cite{hahn+lindquist,eppley,Larry,AHSSS93}. It is comparable to
recent results of Br\"{u}gmann\cite{PWS}: non-headon
black hole evolution through to
significant interaction and merger. But our approach has a novel feature: the
\textit{singularity-excising} character of the computation of generic
encounters which allows ``natural" motion of the black holes through
the computational grid. Singularity excision may be crucial to
carrying out long term simulations predicting gravitational
waveforms through several wave-cycles.

\noindent{{\bf Initial Data:}}
We carry out three binary black hole simulations.
Data is created with {\it spinning} holes, each of mass $m$,
located at $(\pm 5m, \pm m, 0)$, each with Kerr spin parameter $a$. 
The holes are boosted in opposite $\bf \hat x-$directions with speed $c/2$,
representing a grazing collision with impact parameter of $2m$
(and resulting total orbital angular momentum in the $\bf \hat
z-$direction). We distinguish three cases: case (I)--
both holes have $a=0.5m$ opposite to the orbital angular momentum; case (II)--
nonspinning holes $a=0$; case (III)-- both holes have $a=0.5m$
aligned with the total angular momentum.

The total initial ADM mass of each simulation is $2.31m$, which agrees very
well with the estimate given by the special relativistic limit $m_{ADM} = 2
\gamma m$, with $\gamma = (1-.5^2)^{-1/2} = 1.155$.  The
total initial ADM angular momentum ${\bf J}=J \bf \hat z$
is $0.0$,  $1.17m^2$, and $2.34m^2$ for cases I, II, and III respectively
(see \cite{note1}). 

The data setting technique is based on the
boost-invariant Kerr-Schild\cite{KerrSchild} form of the
Kerr black hole metric.  Our {\it Cauchy} formulation
requires first the solution of the initial data problem. As outlined
in \cite{MHS_98,Randy,MHLLMS_00}, superposed boosted Kerr-Schild
data for two single holes produce a conformal background space; the
physical data are solved via a York-conformal approach (solving four
coupled elliptic equations)\cite{YP} on this background. Note that
even when an exact solution of the elliptic equations is known, the error
in the evolved solution will be determined by the inherent evolution-equation
truncation error. Therefore, the accuracy of elliptic solver employed need just
be consistent with this truncation error. For the  discretization used here
($\Delta x = m/4$) the truncation error is of order $5\%$.  The quality
of the data is validated by computing the constraints, normalized to a 
dimensionless quantity by the factor $m^{-2}$.
Analytically the constraints should be zero everywhere. 
In fact with the parameters of the problem, and
with the
current discretization and truncation error, the superposed background
solution is acceptable with no further elliptic problem
solution\cite{MHLLMS_00} (i.e. the 0th order of the elliptic solver). 
However, as we progress to larger and better
resolved evolutions, we will find it mandatory to cycle through the
elliptic solve step\cite{ConsSolve} to obtain satisfactory solution
of the constraints. Figure 1
presents the Hamiltonian constraint for case III,
evaluated at integration timestep $t = 3m$ along the $\bf \hat
x-$axis, together with a time history of the $l_2$ norm (over volume
outside the horizons, and excluding the outer boundary region)
of the Hamiltonian constraint and the similarly normed momentum constraint.   
The late time rise in the momentum constraint in Figure 1 shows the 
beginning of the exponential
mode that appears at about $t=36m$ and ends the simulation.
We have quite good constraint behavior, of order $0.4\%$, 
with peak errors in the Hamiltonian of order $5\%$ until that time.

\noindent{{\bf Evolution Methods:}}
The time-evolutions presented here are done using AGAVE, a code that solves the
Einstein equations in an ADM 3+1 form via finite difference
techniques\cite{Italy}.  A parallel implementation is obtained with
the use of MPI\cite{MPI}, employing the Cactus computational toolkit\cite{cactus} 
solely to aid in this task. AGAVE is a major
revision of the Binary Black Hole Grand Challenge Alliance Cauchy
code\cite{Group,huq+matzner}.
The {\it lapse} function $\alpha$ and {\it shift} vector $\beta^i$
express  coordinate conditions
     which are chosen to allow the black holes to move freely.  For our
simulations, {\it prior} to the time that a single black hole surrounding the
incoming
pair is detected, we use a superposition of functions from boosted black holes:
$
\alpha = \alpha_1 + \alpha_2 -1 \, , \; \beta^i = \beta_1^i +\beta_2^i \, ,$
where these functions are centered with the
current location of the holes, and with the velocity initially obtained from 
Newtonian approximation to the trajectories of the holes and subsequently
inferred from the history of the locations of the
apparent horizons(see below);
{\it after} the detected merger, we use the lapse and shift of a single black
hole with a mass which is the sum of the original bare masses,  and
angular momentum which is the (naive vectorial) sum of the spin and orbital
angular momentum in the original system. (See {\it Discussion}, below.)

The interior of the black holes is excised (Unruh,  quoted in
\cite{thornburg}). We use the apparent horizon surface, locatable at each
time-slice, as a marker for the excision. We utilize a combination of two
different finite difference methods to find the apparent horizon: a direct
solver\cite{Huq}, and a curvature flow method\cite{Deirdre}.  Once the
apparent horizon is located, we  define a {\it mask} function that delineates
the excluded region (interior to the holes) from the computation.  The result
is that we literally evolve two holes moving freely through the computational
domain. That domain is a $161^3$ lattice, corresponding at our resolution to a cube
$(40m)^3$ ($\pm 20m$ in each direction from the centered origin). However,
boundary conditions are set by providing Dirichlet boundary conditions for
$g_{ab}$ and  blending\cite{roberto,luciano} outwards from a sphere of radius $19m$  the
computational solution of $K_{ab}$ to an analytically given (time-dependent)
solution for $K_{ab}$ at the outer boundary sphere. ``Blending" means taking
a linear combination of values from the computed and the analytically given
solution, over a few (here, four) spatial zones, reducing gradients and
second derivatives at the boundary. The analytic blending solution is created
by superposition of boosted holes given by the initial data construction (with
centers and velocities propagated according to the lapse and shift computation), or
after the merger by the final estimated black hole with post merger lapse and shift.

The discretization of the Einstein equations is consistent to second order accuracy. 
On the time scale where instabilities do not play a significant role, the  
convergence rate of this code is  $\approx 1.6$, reduced from $2$ apparently because of extrapolation 
at the excision  boundaries.

\noindent{{\bf Results:}} 
To the current accuracy of the code, cases I-III behave similarly. The total proper area of the
apparent horizon  $A$ for case (I) is shown in Figure 2.  The value of $A$ is particularly interesting
since it provides a measure of the total mass contained in the apparent horizon. For a given
black hole of mass $m$ and spin parameter $a$ its area is $A_{BH}= 4 \pi (R_{+}^2 + a^2)$ 
(with $R_{+}=m+\sqrt{m^2-a^2}$). Since at early
times there is no common apparent horizon the total area is approximately $A \approx A_{BH1} +
A_{BH2} = 2 A_{BH1}$, as the holes merge the total mass enclosed in the common horizon is (roughly)
expected to double, and hence its area would be four times as bigger, ie. for a non-spining final
black hole $A \approx 4 \pi  (2
(2m))^2 \lesssim 4 A_{BH1}$. Therefore, a plot of $A$ vs. time (like the one in figure 2) shows a
considerable `jump' at the time the holes merge  $t\approx3.8m$. 
Additionally, effects of the outer boundary can be clearly seen in figure 2. For a $\pm 10$ grid an abrupt `kink'
is seen at $t\approx10m$ while in the $\pm 20$ grid the `kink' appears at $t\approx 20m$.
At about $t\approx36m$ ($t\approx26m$) apparent instabilities in the $\pm 20m$  ($\pm 10m$) 
grid cause a rapid increase in the
computed horizon size and eventually crash the run. Thus at $t \approx35m$
the solution becomes untrustworthy. While the simulation is free of boundary effects
the coincidence of the measured horizon area values supports confidence in the results.
Figures 3A - 3F track the apparent horizons through the merger for case I.
A single
enveloping black hole appears at $t\approx3.8m$. The horizon
oscillates and grows slightly.

We have in place Cauchy-characteristic
extraction, where the Cauchy solution sampled at some ``large" radius
acts as data for a characteristic evolution to
infinity\cite{Group2,matching} for waveform extraction. We also can compute
the Newman Penrose tensor $\psi_4$, which captures at null infinity
the outgoing radiation.  Additionally, we are developing a perturbative radiation
extraction module.
We are preparing an article explaining how these tools are applied and
illustrating the radiation patterns obtained from these simulations.

\noindent{{\bf Discussion and Future Directions:}}
The simulations reported
here are genuinely, but not excessively, hyperbolic encounters. A Newtonian
estimate gives a free fall velocity of $0.4c$ from infinity, as compared
with the velocity $0.5c$ specified in our initial data. Future work will
concentrate on generic hyperbolic and elliptic orbits.

Ongoing research concerns the late-time stability of the black hole
simulations. We have carried out a number of 1-dimensional simulations, all of
which have longer term stability than this 3-dimensional simulation of
merged holes. We are investigating the behavior of the differencing scheme at
the inner boundary. (The one we use behaves well in the spherical case.) We are
implementing a new outer boundary algorithm which has been shown to be robustly
stable in a linearized version of the code~\cite{szylagyi}.  We are developing
more sophisticated gauges based on elliptic  equations for the lapse and the
shift. These include the minimal distortion and minimal shear
gauges\cite{york+smarr}, and other elliptic gauges\cite{MHS_98,thorne+brady}.
Stable evolution of single black holes is quite sensitive to gauge conditions,
and we anticipate much useful science from future improvement in the
lifetime of our simulations of black hole mergers.

Our gauge and boundary conditions for the final merged black hole naively assume
that all the initial mass  (i.e. $M_{final} = 2m$) and  angular momentum
resides in the final hole: $J_{final} = a_{final}\times M_{final}$.  For cases
I, II, III our gauge takes $a_{final} = (0, 0.25, 0.5)\times M_{final}$.  These
estimates do not take into account the emission of energy and angular momentum
during the dynamics, nor the $\gamma$ factor in the initial mass and angular
momentum. The actual post-collision mass and angular momentum of the residual
hole will be evaluated to further improve the simulations; behavior of the code
is robust under changes in the final assumed mass and spin.

Of extreme interest is the size of the
final apparent horizon.  The {\it total} initial ADM mass leads to horizon area of $4 \pi
(2 \times 2.31m)^2\approx268m^2$. The post-merger numerically computed apparent horizon
area (Figure 2) is about $255m^2$, $5\%$  smaller than this estimate. This measure would 
give a preliminary indication that total energy radiated  in this simulation is about $2.6\%$. 
However, we have
yet to complete a 3-dimensional {\it event} horizon tracker, which will allow a correct
comparison of the initial and final  {\it event} horizon area.

The present work demonstrates the first simulation of binary
black hole systems via the excision of singularities. The 
datasets evolved are not only useful for validation of the
techniques employed here but as valid datasets in an astrophysical sense for the final
``plunge'' of the merger. In this work we: (a) demonstrate well behaved (convergent)
descriptions of the black holes as they evolve; (b) show that apparent
horizon tracking and black hole excision can produce dynamical multi-black
hole spacetimes, with reasonably well controlled errors for a considerable
length of time (long enough for an accurate modeling of the merger phase); and 
(c) demonstrate that relatively unsophisticated gauge functions
$\alpha$ and $\beta$ can lead to physically interesting evolution lifetimes.

This work owes much to the Binary Black Hole Grand Challenge, and 
we thank all the members of that
effort. This work was supported by NSF PHY/ASC 9318152 (ARPA supplemented),  
PHY 9310053, PHY9800722, PHY 9800725 to the University of Texas at Austin; 
PHY 9800973,
PHY 9800970 to Penn State University and 
PHY 9510895, INT 9515257, PHY 9800731 to the University of
Pittsburgh.  RM thanks the Observatoire de Paris, where some of 
this work was carried out, and Los
Alamos National Laboratory for support during the initial stages 
of this research. E.S.
acknowledges support from  DAAD. Computations were carried out 
at the National Center for
Supercomputing Applications, at the Albuquerque High Performance 
Computing Center, and at Los
Alamos National Laboratory

\begin{figure}
\centerline{\epsfxsize=195pt\epsfbox{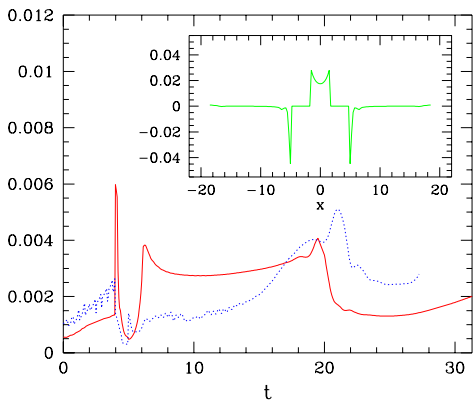}}
\caption{For case I (and grid$\pm 20$), the Hamiltonian and momentum constraints, 
on the domain of outer communication (outside the apparent 
horizon(s) and inside the outer boundary blending zone). We give
the time history of the $l_2$ (rms)  norm of the Hamiltonian (solid line)
and the $l_2$ norm 
over all three components of the momentum constraint 
(dotted line). The momentum $l_2$ is constructed only along 
coordinate lines (all that is available from this computation); 
the Hamiltonian  $l_2$ is computed from the whole volume. 
The sudden change in the errors at $t \approx4m$ occurs when 
a single outer apparent horizon envelops the merging holes.
Also, the drop at $t\approx 20m$ is due to boundary
effects.
The inset shows the  Hamiltonian constraint 
along the $x-$axis at time $t=3m$.}
\label{fig:one}
\end{figure}

\begin{figure}
\centerline{\epsfxsize=205pt\epsfbox{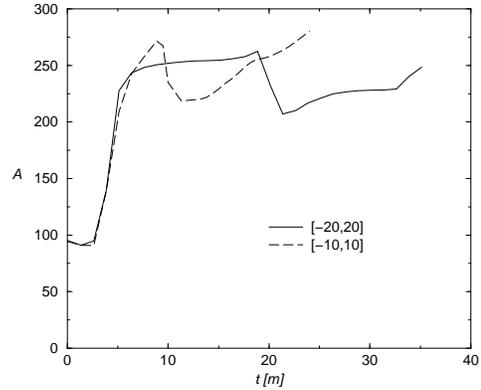}}
\caption{The area of the apparent horizon(s) (transition to a single 
horizon at $t \approx4m$) for case I. For a smaller domain 
($\pm 10m$, dashed line) the simulation runs to $t\approx26m$ and exhibits
strong boundary effects at $t\approx 10m$. 
In the larger ($\pm 20m$) domain (solid line) boundary effects
show at later time, around $20m$. Instabilities cause the measured 
area to rise abruptly at 
$t \approx36m$ and eventually stop the simulation.}
\label{fig:second}
\end{figure}

\begin{figure}
\centerline{\epsfxsize=295pt\epsfbox{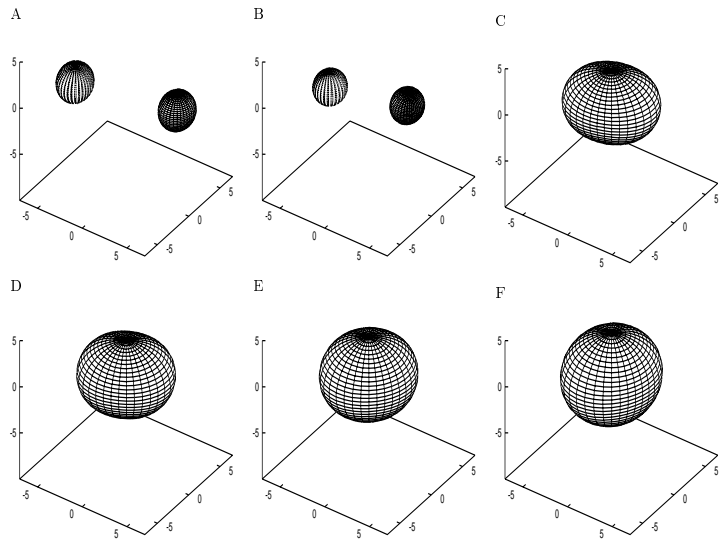}}
\caption{For case I, time history of the horizons. The times 
corresponding to figures 3A-3F are $t=0,$ $ 2.6m,$ $ 5.1m,$ $ 8.8m,$ $ 13.8m,$ $ 18.8m.$ 
These are coordinate plots; the corresponding areas appear in Figure 2.  
After the merger the horizon oscillates through a fraction of a 
cycle.}
\label{fig:three}
\end{figure}


\begin{thebibliography}{99}

\bibitem{LIGO}LIGO: A. Abramovici, {\it et al.},
\textit{Science} {\bf 256}, 325 (1992);
GEO: K. Danzmann and the GEO Team
     {\it Lecture Notes in Physics} {\bf 410} 184-209 (1992);
VIRGO: C. Bradaschia {\it et al.},
\textit{Nucl. Instrum. Meth., Phys. Res. Sect.}, {\bf 289}, 518 (1990);
TAMA:  K. Tsubomo, M.-K. Fujimoto
     and K. Kuroda, \textit{Proceedings of the TAMA International
     Workshop on Gravitational
     Wave Detection} (Universal Academic Press, Tokyo, Japan), (1996);
LISA: \textit{ Proceedings of the First International LISA Symposium}
     \textit{Class. Quantum Grav.}  {\bf 14}, 1397 (1996). 

\bibitem{ADM} R. Arnowitt, S. Deser, and C. Misner, in {\it Gravitation -- An
Introduction to Current Research}, edited by L. Witten
 (Wiley, New York, 1962).

\bibitem{Italy} Richard A. Matzner, in {\it Classical and Quantum Black Holes},
P Fre, V. Gorini., G. Magli, and U. Moschella, editor
s, Institute of Physics Publishing (Bristol 1999).

\bibitem{hahn+lindquist}S. G. Hahn and R. W. Lindquist, 
{\it Annals of Physics (NY)} {\bf 29} 304 (1964).


\bibitem{eppley}K. R. Eppley, {\it Phys. Rev.} {\bf D16} 1609 (1977).



\bibitem{Larry} L. Smarr, in {\it Sources of Gravitational Radiation},
edited by L. Smarr (Cambridge University Press, 1978).

\bibitem{AHSSS93}P. Anninos {\it et. al.},
\textit{Phys. Rev. Lett.} \textbf{71} 2851 (1993);
P. Anninos {\it et. al.}  \textit{Phys. Rev.} \textbf{D52} 2044  (1995);
 P.~Anninos {\it et. al.} {\it Physical Review}, {\bf D52}, 2059-2082 (1995).

\bibitem{PWS} B. Br\"{u}gmann,
\textit{Int. J. Mod. Phys.} {\bf D8}, 85  (1999).


\bibitem{note1} Spatial components of angular momentum (e.g. spin) perpendicular to the
motion transform with one power of $\gamma$ and stay perpendicular to the
motion. The orbital angular momentum ${\bf L} = {\bf r}\times {\bf p}$
also contains one power of $\gamma$ in ${\bf p}$. Hence ${\bf J}$ contains one power of
$\gamma$. See L. Landau and E. Lifshitz, {\it Classical Theory of Fields}
revised second edition, (Pergammon Press, Oxford, 1962), p. 46.



\bibitem{KerrSchild} R.P. Kerr and A. Schild, in
 \textit{Applications of Nonlinear Partial
          Differential Equations in Mathematical Physics}, Proc. of
          Symposia b Applied Math., Vol. XV11, (1965);
         R.P. Kerr and A. Schild, in \textit{Atti del Convegno Sulla Relativita
          Generale: Problemi Dell'Energia E Onde Gravitazionale},
          G. Barbera, ed.(1965).

\bibitem{MHS_98} Richard A.\ Matzner, M. F. \ Huq, and D. \ Shoemaker,
        \textit{Phys. Rev.} {\bf D59}, 024015  (1999).

\bibitem{Randy} R. Correll, PhD Dissertation, The University of Texas (1998).

\bibitem{MHLLMS_00} P. Marronetti {\it et. al.}
        \textit{Phys. Rev.}{\bf D 62}, 024017 (2000).

\bibitem{YP} J.~W. York and  T.~\ Piran
in
\textit{Spacetime and Geometry: The Alfred Schild
Lectures}, Richard~A.\ Matzner and L.~C. Shepley Eds.
University of Texas Press, Austin, Texas. (1982);
G. B. Cook, Ph.D. Dissertation, The University
of North Carolina at Chapel Hill (1990).

\bibitem{ConsSolve} P. Marronetti, Richard A. Matzner. gr-qc/0009044.

\bibitem{MPI} http://www.mpi-forum.org/

\bibitem{cactus}  http://www.cactuscode.org/


\bibitem{Group}
{\it The Binary Black Hole Grand Challenge Alliance}:
G. Cook, et. al.
\textit{Phys. Rev. Lett.} {\bf 80}, 2512  (1998).

\bibitem{huq+matzner}M. Huq and Richard A. Matzner, in preparation (2000).


\bibitem{thornburg} J.\ Thornburg, \textit{Class. Quantum Grav.}, {\bf 4},
1119-1131 (1987).


\bibitem{Huq}M. F. Huq, M. W. Choptuik, and Richard A. Matzner, {\it Phys. Rev.
} in press (2000).

\bibitem{Deirdre} D. Shoemaker, PhD Dissertation, The University of Texas
(1999);
D. Shoemaker, M. Huq and Richard A. Matzner {\it Phys. Rev.} in press (2000).

\bibitem{roberto}R. Gomez, Lecture at Binary Black Hole Grand Challenge
Workshop, Los Alamos NM (October 1997).

\bibitem{luciano}
{\it The Binary Black Hole Grand Challenge Alliance}:
L. Rezzolla, et. al.
\textit{Phys. Rev. Lett.} {\bf 80}, 1812 (1998); 
L. Rezzolla, et. al., {\it Phys. Rev.} {\bf D59} 064001 (1999).



\bibitem{Group2}
N.T. Bishop {\it et. al.} in {\it Black Holes, Gravitational Radiation and the Universe},
eds. Bala Iyer \& Biplab Bhawal (Kluwer, 1998).

\bibitem{matching}N. T. Bishop  {\it et. al.}
 \textit{Phys. Rev. Lett.} {\bf 76}, 4303 (1996).

\bibitem{szylagyi} B. Szilagyi {\it et. al.}. (to appear in {\it Phys. Rev. D.}) (2000).

\bibitem{york+smarr}L. Smarr and J. York, , {\it Phys. Rev.} {\bf D17}
2529 (1978).

\bibitem{thorne+brady}P. Brady, J. Creighton, and K. Thorne, {\it Phys. Rev.}
{\bf D58}
061501 (1998).


\end{thebibliography}
\end{document}